\documentstyle[prl,aps]{revtex}
\begin{document}
In his Comment~\cite{commento}
O'Connell objects that the master equation (ME) I recently derived
in~\cite{art3} (henceforth VME) 
describing the dissipative behavior of a
Brownian particle interacting through collisions with a gas of
Boltzmann particles is not acceptable. To reach this
conclusion he refers to a ME derived
in~\cite{Ford-O'Connell-me} and recently reproposed in
another Comment~\cite{Ford-O'Connell-commento-gao}, which is
essentially the quantum optical ME in the rotating wave approximation
(RWA), but with an explicit expression for the decay rate. 
He then comes to a
comparison with VME inserting in it an harmonic
oscillator (HO) potential.
This modified ME is then compared to
the high temperature limit of~\cite{Ford-O'Connell-me}: 
invariance under the transformation
$x\rightarrow x\cos\theta + p\sin\theta$, $p\rightarrow -x\sin\theta + 
p\cos\theta$ or equivalently $a\rightarrow ae^{-i\theta}$,
$a^{\scriptscriptstyle\dagger}\rightarrow
a^{\scriptscriptstyle\dagger}e^{+i\theta}$ is now lost and a
state having a canonical structure with the HO
Hamiltonian is not a stationary solution. These remarks are correct,
but the problems arising by the changes thus operated on VME
could have been expected from the very
beginning (apart form applying to other published ME, 
e.g.,~\cite{Diosi-Pechukas2}). They are not an appropriate objection to 
VME, even though useful in clarifying a
few issues.
The key result was derived
by Lindblad in a paper~\cite{LindbladQBM} he wrote just after
his work on the generators of completely positive (CP) time 
evolutions. 
In~\cite{LindbladQBM} he considered the Brownian motion of a
quantum HO and showed that in this case one cannot simultaneously
satisfy the three requirements of CP, translational invariance (TI) and
equipartition. This criterion is a meaningful way to compare  
various descriptions of quantum dissipation arising in
different physical communities, as recently done
in~\cite{Tannor-97-00}. 
VME was obtained for the description of a free particle
undergoing Brownian motion,
so that TI 
is a central issue, and plays an essential role
in the derivation. The physically relevant operators are
${\hat {{\sf x}}}$ and ${\hat {{\sf p}}}$, transforming in
the usual way under space translations: a rotation in phase space
has no physical meaning and invariance under this transformation
is not an issue. Obeying CP and TI, the ME
extrapolated to the case of an HO is not expected to lead to the
canonical equilibrium solution: in fact
in~\cite{Pechukas-Grabert} where VME is considered the authors
look for the stationary solution. 
Deviations of the stationary solution from the canonical
form was recently considered in~\cite{Tannor-97-00}, where
these deviations are shown to be a necessary consequence of
TI and vanish in the RWA. The ME
given in~\cite{Ford-O'Connell-me} is obtained
exactly in the RWA, thus granting CP but violating TI, 
as the very authors stressed in an immediately
subsequent paper~\cite{Ford-O'Connell-traslazioni}. This does not mean 
that the derived ME is not acceptable, simply it describes different
physics. The relevant operators are $a$,
$a^{\scriptscriptstyle\dagger}$ and invariance under rotation in phase 
space is now an important issue, correctly taken into account in the
RWA: one then has the canonical equilibrium
solution. 
Note that  the question about the physically
meaningful equilibrium state is not necessarily obvious if one takes
correlations properly into
account~\cite{Tannor-97-00}.
VME 
has all the three \emph{desired} properties: besides CP and
TI it has the expected stationary solution 
$
\exp({-
\beta
{
{\hat {{\sf p}}}^2
/
     2M
}
})
$. 
This exception for the free particle was already mentioned 
in~\cite{LindbladQBM}. 
The main point in~\cite{art3} however is the microphysical derivation,
so that the result is not necessarily meaningful when extrapolated to
another physical context. In the derivation the Brownian limit $m/M
\ll 1$ is essential.  O'Connell before making any comparison takes the
high temperature limit \emph{following Vacchini}: actually I never
take this limit, typical of the HO treatment. In~\cite{art3} Boltzmann
statistics was considered, but this result has already been extended
to quantum statistics, putting into evidence the role played
by the dynamic structure factor~\cite{art4}.
What I considered was the limit of small momentum transfer, analogous
to the Kramers-Moyal expansion leading from the \emph{Master Equation}
to the Fokker-Planck equation.  \par Though the objection raised
against the validity of VME does not hold, the point raised 
leads to a clarification
of some issues not always clearly spelled out. I would also like to recall that for
the unmodified VME the expected canonical distribution is a stationary
solution and to stress that I do not question the general validity of
equipartition. Simply one cannot expect that the extrapolation of a ME
obtained for a specific physical model to another context should work
without flaws.  \par The author thanks Proff.~L.~Lanz and
A.~Barchielli for discussions.  The work was supported by MURST.
\begin{flushleft}
  Bassano~Vacchini \\
  Dipartimento di Fisica and INFN -  Universit\`a di Milano \\
  Via Celoria 16, I-20133 Milan 
\end{flushleft}
  
\end{document}